\documentclass[%
 reprint,
nofootinbib,
 bibnotes,
 amsmath,amssymb,
 aps,
 prl,
]{revtex4-2}

\usepackage{amsmath}
\usepackage{natbib}
\usepackage{graphicx}% Include figure files
\usepackage{dcolumn}% Align table columns on decimal point
\usepackage{bm}% bold math
\usepackage{physics}
\usepackage{color, soul}
\usepackage{appendix}
\usepackage{blkarray}
\usepackage{multirow}
\usepackage{float}
\restylefloat{table}
\usepackage{mathtools}
\usepackage{comment}
\usepackage{lipsum}
\usepackage{textcomp}

\begin{document}
\newcommand\given[1][]{\:#1\vert\:}

\title{Indefinite global time}

\author{Tom Holden-Dye and Sandu Popescu \\ \textit{H. H. Wills Physics Laboratory, University of Bristol, Tyndall Avenue, Bristol BS8 1TL, United Kingdom \\ 
\textit{ } }}

\date{\today}

\begin{abstract}
By studying the set of correlations that are theoretically possible between physical systems without allowing for signalling of information backwards in time, we here identify correlations that can only be achieved if the time ordering between the systems is fundamentally indefinite. These correlations, if they exist in nature, must result from non-classical, non-deterministic time, and so may have relevance for quantum (or post-quantum) gravity, where a definite global time might not exist.
\end{abstract}

\maketitle

Quantum mechanics (QM) is a fundamentally nonlocal theory \cite{bell_1964}, and subsequently also non-deterministic \cite{aharonov_2005, popescu_2014}; it is through non-determinism that nonlocal correlations between entangled systems can be explained without the need for superluminal signalling \cite{eberhard_1978, ghirardi_rimini_weber_1980, eberhard_ross_1989}. As such, we often refer to quantum correlations as being \textit{non-signalling}.
\par
That non-determinism emerges neatly in this way from nonlocality has led to the study of correlations in general probabilistic theories as a means of deepening our understanding of QM \cite{navascues_wunderlich_2009, navascues_pironio_acin_2007, navascues_pironio_acin_2008, dam_2012, buhrman_massar_2005, barrett_linden_massar_pironio_popescu_roberts_2005, barrett_pironio_2005, marcovitch_reznik_vaidman_2007, piani_horodecki_horodecki_horodecki_2006, brassard_buhrman_linden_methot_tapp_unger_2006, brunner_skrzypczyk_2009, forster_winkler_wolf_2009, cavalcanti_salles_scarani_2010, skrzypczyk_brunner_popescu_2009, linden_popescu_short_winter_2007, pawlowski_paterek_kaszlikowski_scarani_winter_zukowski_2009}. An important result of these studies was the realization that perfect nonlocal correlations, stronger than those obtainable with QM, can theoretically exist whilst still remaining non-signalling \cite{quantum_nonlocality_as_an_axiom}. This spurred a search for such correlations, known as `PR-boxes' \cite{dam_2012, buhrman_massar_2005, barrett_linden_massar_pironio_popescu_roberts_2005, barrett_pironio_2005, marcovitch_reznik_vaidman_2007, piani_horodecki_horodecki_horodecki_2006}, or alternatively axioms ruling them out, and presented a new form of non-classicality beyond QM to be explored.
\par
More recently, investigations into non-determinism in causal structures have found that QM correlations with indefinite causal order can be described, which violate Bell-like causal inequalities \cite{oreshkov_costa_brukner_2012, procopio_2015}. An intriguing link has also been made between these correlations and two-time state interpretations of QM \cite{silva_guryanova_short_skrzypczyk_brunner_popescu_2017, guryanova2017exploring}. In making this link, however, it was shown that a distinction can be made between indefinite causal order and indefinite temporal order, and that indefinite causal order can occur without time needing to be indefinite \cite{silva_guryanova_short_skrzypczyk_brunner_popescu_2017}.
\par
This brings us to the question of whether non-determinism, in the form of indefinite temporal correlations, can be found in time. Indeed, the extension of quantum non-determinism to time is of significant foundational interest, and has been proposed in some approaches to quantum gravity \cite{brukner_caslav_2015, hardy, hardy_forposter, rovelli_1991}. In the hope that the study of correlations will once again shed light on the role of non-determinism in Nature,we here utilise a recently developed temporal analogue of the no-signalling condition \cite{brunner_linden_2013} - the \textit{no-backwards-in-time-signalling condition} \cite{guryanova2017exploring} - in order to obtain a characterization of the correlations that are possible within a theory with deterministic definite time. We discover that there exist correlations which are theoretically permitted according to the fundamental principles we preserve here, yet which are simultaneously only possible if globally time is indefinite. This presents a new boundary between classical definite time and non-classical indefinite time to be studied.
\par
\textit{No-backwards-in-time-signalling}.--- The work here represents a continuation of the use of the model-independent \textit{no-backwards-in-time-signalling} (NBTS) thought experiment. Introduced recently by Guryanova \textit{et al.} \cite{guryanova2017exploring}, this thought experiment bares many similarities to nonlocal games such as the CHSH game \cite{clauser_horne_shimony_holt_1969}, but with a novel focus on temporal correlations.
\par
At the heart of the NBTS thought experiment is the \textit{no-backwards-in-time-signalling condition}:
\begin{itemize}
    \item \textbf{The NBTS condition:} In situations where we can establish a clear time ordering, it should be impossible for information to be sent from the future to the past - in other words, \textit{we cannot signal backwards in time} \cite{guryanova2017exploring}. 
\end{itemize}
\par
This is generally taken to be a fairly immutable truth about the natural world, which we should not expect to be violated. Yet, it needn't be violated
in order to provide new insights; initial studies
have proven it to be a much less restrictive condition than it may at first appear. Indeed, there remains significant theoretical possibilities for rich, unexpected phenomena - such as backwards-in-time influences - to occur in probabilistic theories, without requiring backwards-in-time-signalling \cite{guryanova2017exploring}.
\par
The NBTS set-up is as follows. We have two agents, Alice and Bob, who are situated in separate laboratories. They each have some system in their labs. Alice performs a measurement on the system in her lab, and obtains an outcome $a$. She then later receives an input $x$, which has been randomly generated outside of the lab, such that we can assume that the value of $x$ is independent of the value of $a$ or any other part of the system: this is akin to the assumption of measurement-independence in Bell's theorem \cite{bell_1964, guryanova2017exploring}. Alice then uses $x$, and also possibly $a$, to decide some transformation to perform on her system. Finally, she sends her system out of the laboratory. The same procedure also occurs in Bob's lab, where we label the outcome $b$ and the input $y$. For simplicity, we will here take all of these quantities to be single binary bits, i.e., $a,b,x,y\in\{0,1\}$.
\begin{figure}
\includegraphics*[width=0.96\linewidth,clip]{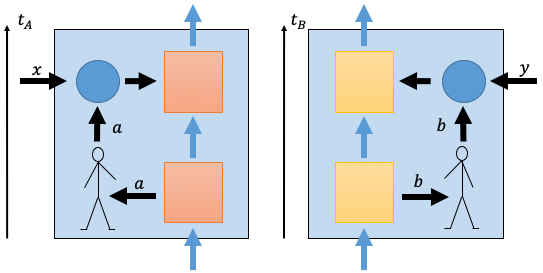}
\caption{The NBTS set-up. An indication of a definite local time ordering within Alice's lab, $t_A$, and within Bob's lab, $t_B$, is shown, highlighting that they both receive their inputs from outside the lab \textit{after} performing their measurements.
}\label{nbtssetup} 
\end{figure}
\par
We assume a definite \textit{local} time ordering within each lab, to ensure that due to the NBTS condition it is impossible for $a$ to depend on $x$, or for $b$ to depend on $y$. We then proceed to consider different relative \textit{global} time orderings between the labs. If one of the experimenters performs their experiment first, then they could in principle send their system to the other lab. This would allow the second experimenter to use the first experimenter’s input and output to produce their own output,
affecting the set of correlations possible between the labs. 
\par 
Considering different global time orderings allows us to make different mathematical statements of the NBTS condition. In the weakest case, the NBTS conditions were derived by Guryanova \textit{et al.} as \cite{guryanova2017exploring}
\begin{equation}\label{indefNBTSconds1}
p_A(a\given x,y) = p_A(a\given y)
\end{equation}
\begin{equation}\label{indefNBTSconds2}
p_B(b\given x,y) = p_B(b\given x),     
\end{equation}
which will be obeyed by all NBTS correlations, no matter the global time ordering. The only restrictions that have been applied are due to the definite local time orderings within each lab, such that $a$ cannot depend on $x$ and $b$ cannot depend on $y$. If no information is known about the global time ordering, then we cannot say for sure that Alice’s system could not pass to Bob, or vice versa - as such, we cannot rule out a dependence of $a$ on $y$, nor of $b$ on $x$. It was noted in \cite{guryanova2017exploring} that the resulting conditions above are identical to the no-signalling conditions in nonlocal games \cite{brunner_linden_2013}, except with the roles of $x$ and $y$ swapped.
\par
In the case that a definite time ordering is known, further constraints can be identified. If Alice’s experiment occurs before Bob’s, $a$ will be produced before $y$ enters the joint system. Our assumption of no-backwards-in-time-signalling then gives us
\begin{equation}\label{defNBTScondsA1}
p_A(a\given x,y) = p_A(a)
\end{equation}
\begin{equation}\label{defNBTScondsA2}
p_B(b\given x,y) = p_B(b\given x).     
\end{equation}
If instead Bob's experiment happens first, we obtain
\begin{equation}\label{defNBTScondsB1}
p_A(a\given x,y) = p_A(a\given y)
\end{equation}
\begin{equation}\label{defNBTScondsB2}
p_B(b\given x,y) = p_B(b).   
\end{equation}
Alternatively, the experiments could occur simultaneously, giving
\begin{equation}\label{defNBTScondsSim1}
p_A(a\given x,y) = p_A(a)
\end{equation}
\begin{equation}\label{defNBTScondsSim2}
p_B(b\given x,y) = p_B(b),     
\end{equation}
the strictest form of the NBTS conditions.
\par
The different sets of correlations, given by the joint probability distributions $p(a,b\given x,y)$, that are achievable under the different time ordering constraints were presented by Guryanova \textit{et al}. Surprisingly, this revealed that there exist correlations which satisfy the NBTS conditions even in the strictest case given in equations (\ref{defNBTScondsSim1}) and (\ref{defNBTScondsSim2}), yet where $p(a,b\given x,y) \neq p(a,b)$; this implies that the future is able to affect the past, without any information being signalled backwards in time \cite{guryanova2017exploring}. For example, we can consider the following PR-like correlation:
\begin{equation}
  p(a,b|x,y) = 
  \begin{cases}
                                   1/2 & \text{if $a \oplus b = x y$} \\
                                   0 & \text{otherwise} .
  \end{cases}
\end{equation}
This satifies all the NBTS conditions so is possible in any of the time ordering cases, but features backwards-in-time influences in the joint probability distribution - $a$ and $b$ jointly depend on the later produced $x$ and $y$.
\par
This raises a pressing question: if such exotic temporal phenomena are already possible when we restrict ourselves to having a definite time ordering, then can \textit{all} correlations consistent with no-backwards-in-time-signalling be obtained in this case - or is more possible when we relax this restriction, and allow for non-determinism in the global time structure? We could imagine using any mixture of all the definite time orderings (Alice before Bob, or Bob before Alice\footnote{Note that we do not need to consider the case of the experiments 
being simultaneous,
where neither experimenter can send their system to the other. The correlations possible in this situation are a subset of those possible when Alice is first, or when Bob is first, so as noted by Guryanova \textit{et al.} it is redundant when considering the limits of what is possible with definite time \cite{guryanova2017exploring}.}) - would this allow us to reproduce all NBTS correlations consistent with equations (\ref{indefNBTSconds1}) and (\ref{indefNBTSconds2})? Or are there some correlations that lie beyond?
\par
It is quite clear to see, when considering the polytopes of definite timing correlations found by Guryanova \textit{et al.} \cite{guryanova2017exploring}, that there is a simple answer to this question. The `guess-your-neighbour's-input' (GYNI) vertices of the NBTS polytope are not present in the polytope formed from the correlations possible when Alice is before Bob, nor in the polytope formed from the correlations possible when Bob is before Alice; we therefore cannot represent the GYNI vertices as a convex combination of any of the vertices from these two definite global time ordering polytopes. Consequently, as any valid definite global time structure should in general take the form of a convex combination of these two time ordering cases, it will be impossible to produce GYNI-type correlations in any situation with strictly definite global time.
\par
Crucially, this means that the correlations of this type cannot possibly be obtained in an NBTS set-up if the global time ordering is confined to being definite - they represent something that must originate from time being truly indefinite. We have found correlations fundamentally and uniquely associated with an indefinite global structure of time, yet which do not require any signalling of information backwards in time.
\par
Specifically, the GYNI vertices are given by
\begin{equation}\label{GYNIvertices}
  p(a,b|x,y) = 
  \begin{cases}
                                   1 & \text{if $a = y \oplus \alpha , b = x \oplus \beta$} \\
                                   0 & \text{otherwise}, 
  \end{cases}
\end{equation}
with $a, b, x, y, \alpha, \beta \in \{ 0,1\}$. In what follows we shall construct general definite timing polytopes in order to explicitly show that NBTS correlations of this form are not possible when there is a definite global structure of time.
\par
\textit{Definite global timing polytope}.--- We can form a polytope out of the collective set of vertices for the $7$-dimensional $A\xrightarrow{}B$ (Alice first) and $A\xleftarrow{}B$ (Bob first) polytopes found in \cite{guryanova2017exploring}, giving an $8$-dimensional polytope with $22$ extremal vertices. Naturally, finding the convex hull formed by all the correlations possible in both of these time ordering cases creates a polytope containing all the possible normalised linear combinations of these correlations. This polytope will therefore contain all of the correlations possible under any such mixture of the two definite time orderings, as required.
\par
Of the $22$ extremal vertices, $12$ are classical deterministic vertices 
\begin{equation}\label{classicaldeterministicvertices}
  p(a,b|x,y) = 
  \begin{cases}
                                   1 & \text{if $a = \mu y \oplus \alpha , b = \nu x \oplus \beta$} \\
                                   0 & \text{otherwise} 
  \end{cases}
\end{equation}
with $\alpha , \beta , \mu , \nu \in \{ 0,1\}$ and $\mu, \nu$ not simultaneously equal to $1$. We also have $8$ PR-like vertices
\begin{equation}
  p(a,b|x,y) = 
  \begin{cases}
                                   1/2 & \text{if $a \oplus b = (x \oplus \gamma)( y \oplus \delta ) \oplus \epsilon$} \\
                                   0 & \text{otherwise} 
  \end{cases}
\end{equation}
with $\gamma , \delta , \epsilon \in \{ 0,1\}$, and finally $2$ linear correlation type vertices
\begin{equation}
  p(a,b|x,y) = 
  \begin{cases}
                                   1/2 & \text{if $a \oplus b = y \oplus x \oplus \alpha$} \\
                                   0 & \text{otherwise} 
  \end{cases}
\end{equation}
with $\alpha \in \{ 0,1\}$. The polytope is $8$-dimensional, consistent with the full NBTS space and NBTS polytope. Previously, only lower dimensional (up to $7$-dimensional) subpolytopes of the NBTS polytope had been obtained \cite{guryanova2017exploring}.
\par
A number of novel facet-defining inequalities, which do not appear in any of the other NBTS polytopes, were found for this polytope: $4$ of the form
\begin{multline}\label{definiteineqs1}
    p_A(a\given y) + p_B(b\given x) \geq \\ p(a,b\given x, y) + p(a,b\given \bar{x}, \bar{y})
\end{multline}
and $4$ more of the form
\begin{multline}\label{definiteineqs2}
    p_A(a\given y) + p_B(b\given x) \\ - [p(a,b\given x, \bar{y}) + p(a,b\given \bar{x}, y)] \leq 1,
\end{multline}
where $x, y, \bar{x}, \bar{y} \in \{ 0,1 \}$ and $x \neq \bar{x}$ and $y \neq \bar{y}$. Note that there would be $4$ times as many inequalities as suggested here if we were to take all the possible binary combinations of $a$ and $b$. However, in defining the NBTS space only probabilities with a certain set $a$ and $b$ are used. The choice of $a$ and $b$ is completely arbitrary, so we have not specified which particular choice we made when presenting the inequalities here.
\par
Although the physical origin of these inequalities is unclear, it is easy to check that they are valid: each inequality is satisfied by all the vertices of the $A\xrightarrow{}B$ and $A\xleftarrow{}B$ polytopes; saturated by vertices from both polytopes; and, most pertinently, violated exclusively by the GYNI vertices.
\par
\textit{Set-q polytope}.--- The polytope found above contains all the correlations possible through any mixture of definite global time orderings. We can also investigate what the polytope for a specific mixture looks like. To do this, we introduce a variable $q$ which denotes the probability that Alice's experiment happens first\footnote{Note that a similar procedure was considered by Guryanova \textit{et al.} but without the inclusion of non-classical correlations, such as the PR-like correlations \cite{guryanova2017exploring}.}. We again ignore the case of the experiments being simultaneous, which then means that Bob's experiment must happen first with probability $(1-q)$. As we vary $q$ from $0$ to $1$ we obtain all possible mixtures of the $A\xrightarrow{}B$ and $A\xleftarrow{}B$ definite global time orderings, and hence all the possible definite global timing subpolytopes.
\par
To construct the set-$q$ polytope (as we will refer to it), its extremal vertices needed to be identified. This was achieved by combining each of the $20$ vertices of the $A\xrightarrow{}B$  polytope with each of the $20$ vertices of the $A\xleftarrow{}B$ polytope using the equation
\begin{multline} \label{combsequationfinal}
    p^{A\xrightarrow{q}B}(a,b\given x,y) = qp^{A\xrightarrow{}B}(a,b\given x,y) \\+ (1-q)p^{A\xleftarrow{}B}(a,b\given x,y),
\end{multline}
where we have denoted the different time orderings as superscript.
\par 
This gives $400$ potential vertices, of which $250$ were then found to be redundant. The set-$q$ polytope formed from the remaining $150$ extremal vertices was found to be $8$-dimensional for all $0 < q < 1$; when $q=1$ it reduces to the $A\xrightarrow{}B$ polytope, and when $q=0$ it reduces to the $A\xleftarrow{}B$ polytope, which are both $7$-dimensional. 
\begin{figure}
\includegraphics*[width=0.96\linewidth,clip]{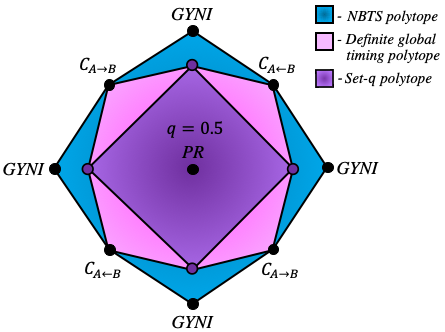}
\caption{The set-$q$ polytope is shown inside the general definite global timing polytope and the full NBTS polytope, in the case that $q=0.5$ . The specific visualisation chosen here clearly shows the existence of the GYNI vertices outside of the definite timing polytope. The classical deterministic vertices for the case of Alice being first, $C_{A\xrightarrow{}B}$, and Bob being first, $C_{A\xleftarrow{}B}$, as well as the PR-like vertices, are also shown. The PR-like vertices appear as if they are inside the body of the polytopes, but they are still extremal vertices - they must extend in some dimension out of the page.
}\label{setqpolytope} 
\end{figure}
\par
In addition to the inequalities defining the general definite timing polytope, the set-$q$ polytope was found to have $4$ novel $q$-dependent inequalities, which can be combined into $2$ pairs (with an upper and lower bound) as
\begin{equation} \label{qdepA}
    -(1-q) \leq p_A(a\given y) - p_A(a\given \bar{y}) \leq (1-q) \\
\end{equation}
and
\begin{equation} \label{qdepB}
    -q \leq p_B(b\given x) - p_B(b\given \bar{x}) \leq q \\,
\end{equation}
again with $x, y, \bar{x}, \bar{y} \in \{ 0,1 \}$, $x \neq \bar{x}$,  and $y \neq \bar{y}$. They are violated by the GYNI vertices, for all possible values of $q$ ($0\leq q \leq 1$).
\par
\textit{Connections to process matrices}.--- Previous studies of no-backwards-in-time-signalling \cite{silva_guryanova_short_skrzypczyk_brunner_popescu_2017, guryanova2017exploring} have discussed the connection between the bipartite correlations achievable here in the NBTS set-up, and those achievable with process matrices - a formalism introduced by Oreshkov, Costa, and Brukner, to study quantum theory with a relaxed causal framework \cite{oreshkov_costa_brukner_2012}. Process matrices can in general allow for non-classical correlations, but not in the NBTS set-up; they can only produce correlations which lie within a classical subpolytope of the definite global timing polytope, which does not contain the GYNI vertices \cite{guryanova2017exploring}. As such, the NBTS set-up and GYNI correlations present a boundary between definite temporal order, yet possible indefinite causal order, and indefinite temporal order.
\par
\textit{Conclusions}.--- The introduction of the NBTS thought experiment presented an analogue of the no-signalling condition for temporal correlations, and a polytope of NBTS correlations was found that was isomorphic to the no-signalling polytope \cite{barrett_linden_massar_pironio_popescu_roberts_2005,guryanova2017exploring}. 
\par
Here, the classification of all correlations possible with a definite global time ordering, including non-classical PR-like correlations, allows another correspondence to be identified; the definite global timing polytope inside the full NBTS polytope bares a striking resemblance to the polytope formed by the set of local correlations inside the no-signalling polytope. We now have a polytope which is a subset of the NBTS polytope with the same dimensionality.
\par
Importantly, this has revealed the existence of correlations which must stem from time being globally indefinite - the GYNI vertices - and presented us with a way to theoretically distinguish between a Universe in which time is globally definite and one in which it is not.
\par
The facets of the definite global timing polytope represent inequalities violated by NBTS correlations that cannot be explained by a classical definite global time ordering, much like the facets of the local no-signalling polytope represent the Bell inequalities which are violated by no-signalling correlations that cannot be explained by classical deterministic hidden variables \cite{bell_1964, barrett_linden_massar_pironio_popescu_roberts_2005}. The GYNI vertices are then the strongest correlations which violate these inequalities and do not allow for backwards-in-time-signalling, in much the same way as the PR vertices of the no-signalling polytope represent the strongest correlations which violate the Bell inequalities and do not allow for superluminal signalling.
\par
We could perhaps ask similar questions about correlations with indefinite global time as were raised about non-classical nonlocal correlations \cite{quantum_nonlocality_as_an_axiom}: Is there something we are missing, that rules these GYNI vertices out? Or could time really be indefinite, as proposed in some theories of quantum gravity without a global time \cite{rovelli_1991}? What connections can be made to quantum theory, and foundational interpretations of it incorporating non-classical temporal structures \cite{aharonov_1964}? These questions are not immediately answered here, but could provide significant motivation for future research.
\bibliography{references.bib}
\end{document}